# Large-scale inference and graph theoretical analysis of gene-regulatory networks in *B. subtilis*

Claire Christensen[1], Anshuman Gupta[2], Costas D. Maranas[2], Réka Albert[1]
[1]*Department of Physics, The Pennsylvania State University, University Park, PA 16802*
[2]*Department of Chemical Engineering, The Pennsylvania State University, University Park, PA 16802*
(e-mail: cpc146@phys.psu.edu)

## Abstract

We present the methods and results of a two-stage modeling process that generates candidate gene-regulatory networks of the bacterium *B.subtilis* from experimentally obtained, yet mathematically underdetermined microchip array data. By employing a computational, linear correlative procedure to generate these networks, and by analyzing the networks from a graph theoretical perspective, we are able to verify the biological viability of our inferred networks, and we demonstrate that our networks' graph theoretical properties are remarkably similar to those of other biological systems. In addition, by comparing our inferred networks to those of a previous, noisier implementation of the linear inference process [17], we are able to identify trends in graph theoretical behavior that occur both in our networks as well as in their perturbed counterparts. These commonalities in behavior at multiple levels of complexity allow us to ascertain the level of complexity to which our process is robust to noise.

# 1. Introduction

Complex networks—systems of multiple interacting parts—can be found in nearly every field of study. The World Wide Web, for example, is a vast network of web pages, interconnected by hyperlinks; social networks are composed of individuals, linked together through their acquaintances; and the North American power grid represents a network of generators and power stations, joined by power lines [1, 38]. With the advent of genome sequencing and high-throughput protein interaction screens, a new realm of complex biological networks has been uncovered. To fully understand the workings of an organism, we must be able to map more than just the organism's genome: in addition, we must understand the networks of metabolic and protein-protein interactions taking place within the organism's cells, and it is critical that we be able to reconstruct and verify the interaction networks through which genes are regulated.

During the last decade, genomics, transcriptiomics and proteomics have produced an incredible quantity of molecular interaction data, contributing to maps of specific cellular networks. Genome-wide protein-protein interaction maps have been constructed for a variety of organisms, including the yeast, *S. cerevisiae* [14,19,21,39], the worm *C. elegans* [28], and the fruit fly *D. melanogaster* [15]. By systematically pairing DNA- binding proteins and their target genes, transcriptional regulatory maps have been constructed for *E. coli* [39] and *S. cerevisiae* [16,27,29]. While these static maps offer significant insights into the remarkable interconnectivity of cellular components, they describe only functional subsets of cellular networks and do not capture the dynamical aspects of gene regulation.

Alternative methods aim to reverse-engineer regulatory networks from gene expression information [18]. Unfortunately, obtaining the data with a low degree of error and with a high degree of completeness is often experimentally impossible; gene expression information lacks clear thresholds, and frequently possesses far fewer measurements than there are genes in the network. In such underdetermined cases, we must turn to mathematical modeling techniques to *infer* candidate regulatory networks of high biological viability. Generally, one of two modeling methods is employed: either a *deterministic model-based* method, in which the expression level of gene X is assumed to affect the expression level of gene Y, or a *stochastic model-based* method, in which the observed expression profiles are assumed to have been selected from a multivariate probability distribution. Because the stochastic model-based method is computationally intensive and underdetermined, here we employ a deterministic model-based approach. Specifically, we identify the regulatory connections among the genes of the bacterium *B. subtilis* through a first-order, linear process [17], and we then perform a multi-scale graph-theoretical analysis of these inferred networks to test the efficacy of the model.

As the extensive work of the last decade demonstrates, a large class of biological graphs shares many features of heterogeneous social and technological networks [1], and its members' topologies are markedly different from those of random graphs. Three dominant graph-theoretical measures define this class of biological graphs. First, members of the class are *scale-free*, meaning that there is a power-law relationship between the number of edges (degree), $k$, adjacent to nodes in the graph, and the fraction, $P(k)$, of nodes with that degree. Second, all graphs in the class exhibit a high average *clustering coefficient* [35], indicating that first neighbors of nodes tend to be highly interconnected. Third, most graphs possess short average path lengths, which is to say that the average distance—i.e. the number of adjacent edges in the shortest path--between any two nodes in the graph is small. In addition, research has revealed that cellular interaction networks exhibit structured patterns of connection at multiple levels of complexity [30,31,34,38], ranging from three-, four-, and five-node network motifs to a global bow-tie structure [5].

Here we compare the characteristics of our inferred networks to the common features of other biological graphs, and by contrasting our networks' properties with those of random graphs, we assess our networks' biological viability. The biological viability of our networks is intrinsically linked not only to their global properties, but also to their subglobal topological characteristics: as the scale of complexity is refined, our networks exhibit a plethora of substructures, revealing a "topological complexity pyramid" [36]. In fact, the regulatory substructures that are revealed by our multi-scale graph-theoretical analysis could provide additional validation of the utility of *synthesizing* large-scale inference and multi-scale graph-theoretical analysis, as these substructures may uncover as-of-yet unknown interactions that can later be verified experimentally. Although this paper reports primarily on the results of a graph-theoretical analysis of inferred gene-regulatory networks in a *specific* organism, the broader goal of this research is to demonstrate that by uniquely integrating experimentation, mathematical modeling, and mathematical analysis, we can uncover both the global characteristics as well as the local interactions inherent to *any* complex system.

## 2. Research Design and Methods

Affymetrix® Gene Chip data, furnished to us by experimental collaborators at Genencor, was used in conjunction with regulatory data from the open literature [3,4,6,7,8,9,10,37] both to construct and to verify candidate gene-regulatory networks. We modeled the dynamics of gene expression as a first-order, linear process where we assumed that the rate-of-change of expression of a particular gene $i$ at time point $t$ is a linear combination of the concentration of all other genes, *i.e.*,

$$\frac{dZ_{it}}{dt} = \frac{Z_{i(t+1)} - Z_{it}}{\Delta t} = \sum_{j=1}^{N} \omega_{ij} \cdot Z_{it} \quad \forall\, i = 1,2,\ldots,N, t = 1,2,\ldots,T-1 \tag{1}$$

where $Z_{it}$ is the expression level of gene $i$ at time point $t$, $\omega_{ij}$ is the regulatory impact of gene $j$ on gene $i$, $N$ is the total number of genes, $T$ is the total number of time points and $\Delta t$ is the sampling interval. This multiple-regression-based linear model has been widely used for inferring gene-regulatory relationships from microarray time series data [12,13,17,20,41]. The relative simplicity of Equation 1 allows both for explicitly correlating changes in the expression of given genes with the expression level of other genes in the network, as well as for determining the effects of these changes on other genes. We inferred the regulatory coefficients $\omega_{ij}$ by solving the system of linear equations that results when experimentally obtained $Z_{it}$ values are plugged in Equation

(1) [17]. The resulting $\omega_{ij}$ values then determined whether gene $j$ up-regulates ($\omega_{ij} > 0$) or down-regulates ($\omega_{ij} < 0$) gene $i$. Using these inferred regulatory coefficients, the gene-regulatory networks were defined [17] such that each gene corresponds to a node and such that a directed edge originates from each "regulator" gene $j$ and terminates on a "regulated" gene $i$ for which $\omega_{ij} \neq 0$. A permutation-based significance analysis was also carried out on the inferred connections: a confidence level was assigned to each inferred connection by comparing the likelihood of inferring a regulatory coefficient of a given magnitude from real versus label-randomized data [17]. Label-randomization was achieved by permuting the rows (gene names) and/or columns (time points) of the expression matrix, and the confidence level for a given regulatory coefficient was found from

$$c_{ij}(\omega^*) = \frac{N^a(\omega_{ij} \geq \omega^*)}{N^a(\omega_{ij} \geq \omega^*) + N^r(\omega_{ij} \geq \omega^*)}, \tag{2}$$

where $N^a(\omega_{ij} \geq \omega^*)$ and $N^r(\omega_{ij} \geq \omega^*)$ are the number of connections inferred from the actual and randomized data sets, respectively (note that because of the discrete nature of the data, $N^a$ and $N^r$ can have a maximum value of $T$-1) [17]. For example, if for a given gene pair, no connections with regulatory coefficients greater than or equal to some prescribed value, $\omega^*$, were inferred from the randomized data, then all connections for that gene pair with regulatory coefficients greater than or equal to the prescribed value in the real data were assigned a confidence level of 100%. We then defined a confidence threshold, and disregarded all edges whose confidence was lower than this threshold. As a result some genes became isolated after losing all edges to which they were previously adjacent; such isolated genes were subsequently excluded from the networks, and only the remaining, connected core of each original network was examined. Therefore, only a subset of the 747 experimentally-represented genes appeared in each inferred network.

Because transcription factors are expressed at much lower levels than enzyme-encoding genes, very few of the 747 experimentally-represented genes encode transcriptional regulators, and at high confidence, the inferred networks contain only a handful of transcription factors. Thus, the edges of our networks mostly correspond to indirect regulatory relationships (probably mediated by transcription factors that were not incorporated in the present analysis).

The data sets correspond to a *(i) B. subtilis* cradle-to-grave experiment with 20 time points; and *(ii) B. subtilis* amino acid pulse experiment with 9 time points. Three networks were generated for each data set at confidence levels of 70%, 80%, and 90%, respectively. Though the extraction method used to generate both of these groups of networks was identical, the two groups represent different experimental conditions. This difference is underscored by the fact that there is only a 1% intersection in the edge sets of the two network groups, even though their node sets differ by only 25%. Luscombe *et al.* [29] have noted a similar trend in the transcriptional regulatory network of *S. cerevisae:* different environmental conditions activate distinct subsets of the full regulatory network, subsets that are overlapping in their nodes (genes) but (largely) non-overlapping in their edges (regulatory interactions).

The networks referenced in [17] were constructed using the same mathematical model and from the same raw microarray data. These arrays exhibit a small but non-disregardable degeneracy of microarray channels: for 21 of the 747 genes examined, between two and five channels of the microarray corresponded to the same gene. In reference [17] all channels were treated as unique genes; here we have collapsed the expression data of each of the 21 genes that were redundantly represented in [17], into a consensus expression profile by averaging the expression profiles of the over-represented genes and their respective repeats at each time step. This profile averaging is justified by the fact that the time course readings of probes corresponding to the same gene are highly similar. Because the networks in [17], formed in the absence of profile averaging, were inferred assuming a gene space with approximately 3% more genes than were represented experimentally, we regard these networks as *perturbed* versions of our inferred networks, and we use them to test the robustness of our inference process by comparing their graph-theoretical properties to those of our networks. The numbers of (non-isolated) nodes and edges for each network (unperturbed and perturbed) at each confidence level are reported in Table 1. The inferred networks are available, upon request, from the corresponding author.

A suite of preexisting *Python* code [11] was employed and augmented for the graph-theoretical analysis of the *B. subtilis* data sets. In addition, the *Mfinder* software package, available from http://www.weizmann.ac.il/mcb/UriAlon/, was used to isolate and enumerate motifs involving between three and five nodes. This software package determines the concentration of each motif, i.e. the number of occurrences of the particular motif of *n* nodes and *m* edges, divided by the total number of motifs of *n* nodes and *m* edges) in a candidate network. Then this concentration is compared to the concentration of the same motif found in a large ensemble of randomized networks that preserve the degree of each node in the candidate network, while establishing connections randomly. The presence of a motif is deemed statistically significant if its concentration in the candidate network had a Z-score greater than 2—i.e. if $\frac{C_i - \mu_i}{\sigma_i} > 2$, where $C_i$ is the concentration of motif *i* in the candidate network, and $\mu_i$ and $\sigma_i$ are the mean and standard deviation in the concentrations of motif *i* in the randomized networks.

## 3.   Results

## A. The global topology of *B. subtilis* networks

### *1. Degree Distribution*

Because the edges in gene-regulatory networks represent the propagation of a chemical signal from one gene to another, the edges in gene-regulatory networks must be *directed*. Edges that point away from nodes contribute to their *out-degree*, and those edges that point toward nodes contribute to their *in-degree*. In the network as a whole, there are necessarily as many outgoing edges as there are incoming edges, but the distribution of outgoing and incoming edges can be markedly different [16,27,38]. Here we study the cumulative in and out degree distributions which give the fraction of nodes with in/out degree $K^{in/out} > k$. Degree distributions that conform mathematically to a power law, $P(k) \sim k^{-\gamma}$, have a cumulative degree distribution,

$$P(K > k) \sim k^{-\gamma+1};  \qquad (3)$$

we look for this feature in our inferred networks.

At all confidence levels, the candidate *B. subtilis* networks, both unperturbed and widely perturbed, exhibit heterogeneous topologies (see Table 2), a characteristic that has been observed in biological networks [16]. As shown in Figure 1, while the total and out-degree distributions are scale-free (with exponential tails), the in-degree is better described by an exponential distribution. This duality of more variable out-degree than in-degree is shared by *E. coli* and yeast transcription networks [16,27]. From Table 2 and Figure 1, it is evident that the out-degree distribution exponents fall very close to 2. A degree exponent approximately equal to 2 has been observed in a variety of cellular networks [16,22,27], and seems to be a characteristic that sets biological networks apart from other networks whose degree distribution exponent is closer to 3 [2].

The heterogeneity in connectivity exhibited by our networks implies that each network possesses a handful of *hub* nodes that have far more neighbors than do other nodes in the graph. The presence of these hubs underscores the fact that the regulatory processes within the inferred networks do not happen in isolation, but rather that they are strongly interwoven with the functioning of other processes within the network; the hubs act as direct links among these processes. The hubs (nodes whose degree is at least 25% of the highest degree in the network) of both sets of networks are genes encoding enzymes (see Table 3). The deviation of this result from the expectation of hub genes encoding transcription factors is explained by the fact that transcription factors are vastly underrepresented in our networks. Indeed, the low expression level of transcription-factor- encoding genes biases our inference method against identifying direct transcriptional regulatory events, and consequently the interactions we infer are likely indirect relationships that are separated by intermediate interactions. While this trend may be an artifact of the inference process, it might also indicate that the most active proteins in the networks indirectly influence the regulation of other genes. We do not observe absolute conservation in the high-degree hub nodes between the unperturbed networks and their perturbed counterparts, however between 30% and 40% of the hub nodes of the perturbed versions of the cradle-to-grave and amino acid pulse networks are similar in regulatory function to hubs of the respective unperturbed networks.

### *2. Clustering Coefficient*

While random graphs have clustering coefficients that scale inversely with the size of the graph,

$$C_{rand} = \frac{\langle k \rangle}{N} \quad [1], \qquad (4)$$

the clustering coefficients of most real networks are unaffected by their sizes [1]. We therefore expect biological networks to possess clustering coefficients that are substantially higher than the clustering coefficients of equivalently large random graphs. For the *B. subtilis* networks, this is precisely what we see. Each network, whether perturbed or unperturbed, has a clustering coefficient two to three times larger than the averaged value for a sampling of random graphs with the same number of nodes and edges (but for which the degree distribution has not been preserved), and roughly an order of magnitude larger than the clustering coefficient of a degree-distribution preserving random graph. Moreover, the ratio of the average clustering coefficient to the average degree for each network is remarkably stable and graph-size independent, differing by less than 10% across confidence intervals, for both the cradle-to-grave and amino acid pulse networks.

### *3. Distances*

Because of edge directionality, the existence of a path from node *i* to node *j*, with path length $d_{ij}$, does not preclude the absence of a path from node *j* to node *i*, making the distance from node *j* to node *i*, $d_{ji}$, infinite. To accommodate these infinite distances, we have calculated the efficiency— the average of inverse distances—of each network:

$$eff = \left\langle \frac{1}{d_{ij}} \right\rangle, \quad \text{where } d_{ij} \text{ is the distance between nodes } i \text{ and } j, \text{ and where the} \tag{5}$$

average is over all node pairs [25,26].

While the efficiency of each network is fairly high—i.e. while each network's average path length is small—it is not markedly different from the efficiency of a random graph that preserves the original network's number of nodes and (directed) edges. It is important to note, however, that while the random graphs support *at least three-quarters* of the total possible paths among nodes, our inferred networks are composed of *less than 25%* of all possible paths. The sparseness of paths in the inferred networks gives a valuable clue as to the structure and function of these graphs, as it suggests not only that all nodes are not connected to all other nodes, but moreover that the directionality of connection is likely crucial to the proper functioning of the network as a whole: signals from one part of the network to another are very specifically and efficiently routed from one gene to another.

### *4. Error/attack resilience and modularity*

To focus on the structural connectivity of the inferred networks, we next consider their undirected scaffold. While this simplification does obscure information about *specific* regulatory interactions, it still provides insight into the networks' susceptibility to node loss. Originally the *undirected* versions of our inferred networks are connected. At all confidence levels, random removal of nodes causes the size of the largest remaining cluster—i.e. the largest group of nodes that can be connected by a continuous path-- to decrease approximately linearly. In fact, the largest remaining cluster typically retains all of its nodes except for the node that was randomly removed. However, removal of the highest degree nodes (see Table 3 for their identities and functions) results in a rapid decrease in the size of the largest remaining cluster within the networks (Figure 2). This pronounced difference in behavior as a function of node removal strategy is another mark of the high degree of heterogeneity exhibited by the candidate networks [1,2], and indicates the important role of the few very high-degree hub nodes (nodes whose degree is at least 25% of the highest degree in the network), in connecting the other nodes in the network. While the hub nodes are responsible for linking large portions of the networks to one another, they are not the sole source of connectivity within each network, as removal of a network's highest degree hubs does not cause the immediate disintegration of the network. In fact, approximately 10% of the nodes of each network must be removed (or all nodes with degree greater than 8 in the amino acid pulse network, and with degree greater than 10 in the cradle-to-grave network) before the network becomes fully disconnected, and this relative resilience to targeted attack implies the existence of additional linking pathways (that do not include the hubs) among the neighbors of the highest degree hubs. In fact, if we investigate nodes (intermediate hubs) whose degree is between 10% and 25% of the maximum degree in the network, whose clustering coefficient is greater than or equal to .4, and that are also neighbors of at least one hub node, and if we look for edges that connect these nodes' neighborhoods without starting or terminating on a hub, we find that on average, there are 3 (5) such edges per intermediate hub in the amino acid pulse (cradle-to-grave) network. We point out that these additional pathways are not necessarily shortest paths among node pairs; they simply function to tie different local neighborhoods together. Due to the relative abundance of these pathways, our networks show no strong evidence of *modularity* in its graph-theoretical sense; i.e. nodes do not appear in clusters that are connected to other clusters through *only one or two* individual edges.

## B. Global topology and subglobal order

Taken in conjunction, the degree distributions, clustering coefficients, path lengths, error/attack resilience plots, and lack of modularity for the *B. subtilis* networks provide valuable hints as to the nature of underlying structure and order within these networks. For example, the existence of nodes of very high degree, whose removal results in the disintegration of the networks, combined with the fact that all nodes lie within a few edges of all other nodes, suggests that the networks' high-degree nodes are *functional* hubs, routing chemical signals quickly and efficiently from one part of the network to other, otherwise distant, network components. The inferred networks' high clustering coefficients, corroborated by the networks' surprisingly high resilience to attack, suggests both that the networks are composed of many small clusters of nodes that are heavily interlinked, and that these small, strongly-linked neighborhoods are bound to one another both via the hub nodes and via pathways that *do not* traverse the hub nodes. To test the accuracy of these hypotheses, we turn to increasingly local topological scales.

*1. Strongly-connected components, in components, out components, and tendrils: the first level of subglobal order*

Nodes in a directed network can be classified according to the ways in which they are bound to other nodes [32]. All nodes that can both reach and be reached from one another form a *strongly-connected component*; those nodes that can reach the strongly-connected component, but that cannot be reached from the strongly-connected component form the *in component* of a network, and those nodes that are connected in a converse manner form the *out component* of the network. Finally, those nodes that can neither reach, nor be reached from, the strongly-connected component of a network form a group of *tendrils*. At highest confidence, both networks contain a strongly-connected component containing more than 10% of the total number of nodes in the networks. Furthermore, between one-third and one-half of these nodes are also among the networks' highest-degree nodes. 65% (87%) of nodes lie in the cradle-to-grave (amino acid pulse) network's out component, while 7.2% (0%) of the nodes belong to the in component. The cradle-to-grave network also has a sizeable tendril component (13.4% of the network), though tendrils are all but absent from the amino acid pulse network.

The strongly-connected component forms a dense region of intersecting paths; through the strongly-connected component, nodes are able to connect to other nodes in the network. Removal of the strongly-connected component causes the isolation of more than one-third of both networks' nodes. Because such a large fraction of the nodes of the networks lose connectivity when the strongly-connected component is removed, it is difficult to compare average path lengths before and after the strongly-connected component's removal; therefore we calculate the ratio of the *relative change in total efficiency and the relative change in total number of paths,* in going from a network to its counterpart without a strongly-connected component, i.e.

$$\frac{1 - \sum_{i'j'} \frac{1}{d_{i'j'}} \Big/ \sum_{ij} \frac{1}{d_{ij}}}{1 - \frac{(N - N_{scc})(N - N_{scc} - 1)}{N(N-1)}}, \qquad (6)$$

where $N$ is the number of nodes in the original graph, and $N_{scc}$ is the number of nodes in the strongly-connected component. The change ratio for the unperturbed cradle-to-grave network at highest confidence is 17.1%, indicating that the number of paths changes more than does path length. Taken in conjunction with the large numbers of nodes that become disconnected following the removal of the strongly-connected component, this information corroborates the idea that the strongly-connected component maintains a dense mesh of paths among a large fraction of the network's nodes. Furthermore, the prevalence of high-degree nodes in the strongly-connected component supports the idea that one means of linking the small, tightly-knit neighborhoods of the *B. subtilis* networks is through paths involving the networks' hubs. Similar results are obtained both for the perturbed cradle-to-grave network at highest confidence, as well as for the unperturbed and perturbed amino acid pulse networks at highest confidence.

One expects that the nodes of each of these network components are involved in processes that are functionally similar; indeed, in mammalian signal transduction networks, nodes of the strongly-connected component are responsible for central signal processing, while the nodes of the in and out components work, respectively, to guide input from ligand-receptor binding and to propagate output from the strongly-connected component to transcription of target genes and phenotypic changes [32]. We also see hints of this type of functional grouping in our networks: most strikingly, the strongly-connected component of the amino acid pulse network contains nearly twice the concentration of transferases and dehydrogenases as does this network's out component. Transferases systematically move chemical groups from one compound to another, while dehydrogenases participate in the oxidation reactions that convert small organic molecules into energy; the high concentration of transferases and dehydrogenases in the strongly-connected component of the amino acid pulse network suggests that the bulk of amino acid metabolism is carried out by the genes comprising the strongly-connected component of this network, since amino acids must first be chemically decomposed by transferases and then dehydrogenated in order to be fully metabolized by the organism.

*2. Unperturbed networks and perturbed networks: a stricter comparison of their global topologies*

As was mentioned previously, we studied our inference method's resilience to perturbation by comparing the graph-theoretical properties of our networks to the graph-theoretical properties of networks inferred under the assumption that 21 redundant channels of the microarray represented unique genes (we refer to these networks as "perturbed"). While the number of genes initially considered in the inference process was similar (747 vs. 695 ), only non-isolated nodes are considered part of the inferred network (see Sec. 2, Research Design and Methods); thus the overlap between the number of genes in the perturbed and unperturbed networks is non-trivial. Interestingly, while (at highest confidence) the perturbed networks were less than 10% larger than their unperturbed counterparts, the overlap in genes common to an unperturbed network and its corresponding perturbed network ranged from a mere 45.6% to 59%.

At first glance, this low level of conservation would seem to suggest that the inference process is highly sensitive to perturbation; however, if the nodes of a perturbed network are chosen by randomly drawing genes from the pool of 747 genes, over $10^5$ iterations of this process, 100% of randomly-drawn networks exhibit node overlap with the unperturbed networks that is non-trivially lower than the overlap in our perturbed networks. For this reason, we conclude that the node overlap observed in our networks exhibits robustness in terms of node selection. However, the process seems to be hypersensitive to noise in establishing regulatory connections: regardless of the network and confidence level, at best only 1.1% of an unperturbed network's edges appear in the perturbed graph. This lack of robustness in regulatory connection conservation is perhaps not surprising: the linear inference method used to create candidate networks predicts a *minimal set* of *indirect* regulatory interactions; therefore, an edge in one inferred network might correspond to a multi-edge path, or could be rejected as being low confidence, in a network inferred under slightly different conditions. Indeed, we see some evidence of edge-to-path conversion in our inferred networks: 10% of edges in the amino acid pulse network are found as multi-edge paths in the network's perturbed counterpart, and 29% of edges in the cradle-to-grave network are found as multi-edge paths in this network's perturbed counterpart. Furthermore, as an analysis of the subglobal topologies of the unperturbed and perturbed networks reveals, the inference process is highly effective in consistently producing patterns of subglobal (and global) structure and order that are common both to the unperturbed and perturbed networks.

*3. Commonalities in structure and function at the most basic level of complexity*

**Network motifs**: The pioneering work of the Alon group in 2002 [34,38] demonstrated that nearly all complex networks possess subglobal patterns of connection that recur with a greater statistical significance than would be found in degree preserving random graphs —i.e. in graphs for which both the degree of each node, as well as the number of nodes with a specific in- or out-degree are preserved. These *network motifs* have garnered great interest in systems biology, as it has since become evident that biological networks with similar functions possess the same motifs. Transcriptional regulatory networks, for example, have been shown to contain, predominantly, *feed-forward loops, bi-fans,* and *single-input modules* [34,38], whereas signal transduction networks also show evidence of *feed-back loops* [32, 38]. A diverse array of motifs, involving between three and five nodes, emerges in the *B. subtilis* networks; moreover the most abundantly-observed patterns within our simulated networks have also been identified in other gene-regulatory networks[30,31,34,38,43]. Among three nodes, for example, the dominant connection patterns in our networks involve feed-forward loops; as the scale of complexity is increased from three nodes to five nodes, the most abundant motifs in the inferred networks are extended feed-forward loop structures (Figure 3). Moreover, similarities in structure between the most statistically significant motifs of the unperturbed networks and their perturbed counterparts demonstrates the ability of our network inference method to consistently construct networks that are built from simple, nonrandom regulatory structures. Although some differences in motif identity and abundance exist between the unperturbed and perturbed networks, the qualitative commonalities in motif structure between the two sets of networks suggest a robustness in the basic network patterning that arises as a result of the inference process.

In addition to the highest-abundance feed-forward-loop-based structures, 30 other three-, four-, and five-node motifs were also found over a range of high statistical significances in at least one of the two unperturbed candidate networks; after perturbation, 18 motifs of high statistical significance were found in the candidate networks (see Appendix for a list of motifs that were found in two or more networks at highest confidence, with Z-score>2). Here we focus on motifs that appeared with high significance (Z-score>2) in both the cradle-to-grave and amino acid pulse networks (unperturbed and perturbed) at highest confidence.

In order to better quantify the extent to which high-confidence motif inference is robust, we examined whether or not the highest confidence motifs in an unperturbed network were identically reproduced in the corresponding perturbed network; that is, we determined whether the motifs with greatest abundance in an unperturbed network were composed of *exactly* the same genes as the most abundant motifs of the same size in the perturbed network. With the exception of the five-node motif in the cradle-to-grave networks, which did exhibit node conservation, there appears to be almost no complete node-wise reproduction of motifs. Nonetheless, there is evidence of *partial* node-wise motif reproduction in going from an unperturbed network to its perturbed counterpart, as between 1% and 5% of high-confidence motifs contain at least one gene common to both the unperturbed and perturbed networks.

While the lack of complete node-wise conservation of motifs within our inferred networks is most likely a result of the inference process, there is strong evidence that complete node-wise conservation of regulatory motifs between networks corresponding to different environmental conditions is actually *unlikely* [29,32,40]. In light of these arguments and the fact that there is scant edge conservation between our unperturbed and perturbed networks, it is not surprising that there is only a small degree of node-wise conservation among the motifs of greatest abundance in these networks. Because of this expected variability in motif composition, it has been suggested that node-wise conservation is of secondary importance to the ability of motifs both to predict and to be predicted from the global topology and function(s) of a particular network [40]. Given this reasoning, we found it more meaningful to examine the conservation of information-propagating *functions*— both of the highest abundance motifs, as well as of the nodes that occurred most often in these motifs—between unperturbed and perturbed versions of the inferred networks. We argue that although the conservation of specific nodes in specific motifs is not necessarily expected, if our inference process is robust to noise, and if the inferred graphs reflect a biological reality, there should be conserved, dominant features in the patterns of signal propagation (where signal directionality is signified by the direction of an edge between two nodes) within the networks that are neither artifacts of the inference process, nor byproducts of noise, but that are, in fact, responses to environmental stimuli.

**Node roles: conservation of information-propagating function**: In verifying the hypothesis put forth in the previous section, we observe that a specific node, appearing in more than one of a given type of motif, may adopt different *roles* from one motif to the next. For example, a node that appears in two feed-forward loops may appear as the node with two outgoing edges in one motif, but it may then become the node with two incoming edges in the second motif. This further suggests that absolute conservation of motifs may not be necessary—that a variety of nodes can "stand in" for a given node within a single network, or from one network to another. Such "network rewiring" has been noted in other gene regulatory networks [29].

To obtain a clearer picture of the specific functions associated with these multiply-incident nodes, and, subsequently, to uncover the functional role of the motifs in which the nodes are found, we define five node *roles* (Figure 4) based on directionality and abundance of edges incident to a node, and we then categorize the motifs according to the roles of the nodes from which they are composed. Thus, *sinks* are nodes with incoming edges only; *sources* are nodes with outgoing edges only; *relays* have equal numbers of incoming and outgoing edges; *branches* incorporate one incoming edge and multiple outgoing edges; and *integrators* merge multiple incoming edges with one outgoing edge. If we think of node roles as being linked to the propagation of chemical signals throughout the network, sinks then represent signals' final destinations, while sources indicate those signals' origins. The presence of a relay likely increases the efficiency of signal propagation by creating a path between two nodes that would otherwise be separated by, for example, the spatial distance of the genes' products within the cell. Alternatively, a relay's presence might act to delay a signal's arrival at a sink node, in the case that other signals must reach the sink node before the arrival of the delayed signal. Finally, branches likely work to disseminate a signal—to branch it out-- and integrators, likely concentrate and synthesize a number of incoming chemical messages to integrate them into one downstream signal.

It is important to realize that a single node can be part of more than one motif, and further, that it may have different roles in different motifs. Within a single candidate network, a handful of nodes appear statistically more frequently in significant network motifs than do the other nodes in the network, and these "high-frequency" nodes may adopt different roles in different motifs. The five high-frequency nodes that appear in the widest variety of motifs are listed in Table 4. When high-frequency nodes' roles are tallied across motifs, each shows a preference toward a specific role. Of the 26 high-frequency nodes in the amino acid pulse network, 23 are sinks; the high-frequency nodes of the perturbed amino acid pulse network also appear predominantly as sinks. In the cradle-to-grave network, on the other hand, out of 21 high-frequency nodes, nine— including the five nodes that participate in the widest variety of motifs— act predominantly as branches. This dominance of branching behavior is also present in the high-frequency nodes of the perturbed cradle-to-grave network.

To truly understand the function of a given motif in the network as a whole, that motif's *direct* connections to other nodes in the network must be taken into consideration. Consequently we aim for a classification of motifs according to their overall functions in a gene-regulatory network: in particular, we can determine whether a given motif is primarily the target of chemical signals from other parts of the network, whether it acts as the origin of chemical signals, or whether it works to transform and then route signals from one part of the network to another.

**Functional classification of motifs based on node roles**: The high-significance motifs in the amino acid pulse and cradle-to-grave networks can be divided into classes based on the roles of their constituent nodes. In the amino acid pulse network, all high-significance motifs contain at least one sink, and 11 of these 14 motifs are comprised of nodes with at least three of the following four roles: branch, integrator, source, or sink. In the cradle-to-grave network, however, only 11 of the 17 high-significance motifs possess a sink, and in 12 of the 17 high-significance motifs, the nodes--one of which is always a branch or integrator—have only two roles.

If the sink in a high-significance motif is truly a sink, that is, if the motif's sink node has no outgoing edges, the motif can be said to act as a signal repository—whether the signal originates in the motif, itself, or whether it enters the motif from another part of the network, the signal will terminate in the motif's sink. Similarly, if the source in a motif is a *true* source—i.e. if it has only outgoing edges—the motif can be thought of as the source of a signal. While we did not find any high-frequency nodes of either network that were true sources, 23 of the 26 high-frequency genes in the amino acid pulse network are true sinks, and, because every high-significance motif in this network involves at least one high-frequency gene that is a true sink, this implies that all high-significance motifs of the amino acid pulse network are signal repositories. The cradle-to-grave network shows a marked difference in structure from the amino acid pulse network, as none of its high-frequency nodes are true sinks.

This difference in behavior between the high-frequency nodes in the statistically-significant motifs of one network and their counterparts in the other network suggests a difference in overall function of these networks' motifs. We propose that the motifs of the cradle-to-grave network act primarily as signal transformers and disseminators showing little sign of signal instigation or termination, because the bacteria in this particular set were treated with an *unlimited and nutrient-balanced* solution—there was no specific nutrient that dominated the nutrient cocktail on which the bacteria fed, and, therefore, there was no one gene-regulatory system that appeared to be more active than other systems. The high-frequency nodes of this network have a variety of gene functions, indicating that there is no one particular task with which these genes are concerned. Thus, in the cradle-to-grave network, endogenous signals flow among all of the regulatory systems in the bacteria, aided in their transmission by the branching nodes of the high-frequency motifs evident in this data set [32].

In the amino acid pulse network, however, which had been treated with a nutrient cocktail composed of only *one type of nutrient* (namely, amino acids), we can isolate the terminal points in many of the gene-regulatory pathways, which suggests that exogenous information affects different parts of the organisms' gene-regulatory networks in different ways: this signal-concentrating effect which was absent in the absence of biased external stimuli implies a cause-and-effect relationship—it implies *reaction*. Indeed, 16 of the high-frequency nodes in the amino acid pulse network represent genes involved in amino acid metabolism, further suggesting that their regulation is a response to the exogenous signal. This may be the most direct evidence that our inferred regulatory networks accurately model gene regulation in *B. subtilis*: we are "seeing" the metabolism of the amino acid cocktail embedded in the topology of the amino acid pulse network.

Similar behavior is noted in the perturbed amino acid pulse network, though there are fewer sinks overall.

**Regulatory pathways**: The idea that node (and, therefore, motif) role is largely determined by a network's response to endogenous and exogenous signal propagation is further supported by the fact that many of the highest frequency genes within a given network can be grouped according to the metabolic or regulatory function(s) associated with them. Between one-third and one-half of the high-frequency genes in each network (unperturbed and perturbed) share a regulatory pathway, according to the *KEGG2* archive [26]. While the high-frequency genes of the unperturbed networks are not found as high-frequency genes of the perturbed networks, all of the regulatory pathways—three in each network—are represented in the perturbed networks. These regulatory pathways, the high-frequency genes that compose them, and their primary regulatory function are given in Table 5. We point out that the networks' high-frequency genes are not necessarily the hubs of the networks. Although it is true that the greatest aggregation of motifs will occur around hubs [40], it is not true that motifs of highest statistical significance will necessarily be linked to hub nodes; therefore, it is possible that the nodes identified as high-frequency genes will not be hubs. In light of the fact that gene conservation after perturbation is less than 60% in both the amino acid pulse and cradle-to-grave networks at highest confidence, it is perhaps not surprising that very few of the high-frequency genes in the unperturbed networks are also high-frequency genes in the perturbed networks (Table 5). Nonetheless, we do observe that approximately one-third of the high-frequency genes in the unperturbed networks have a *similar* counterpart in the corresponding perturbed networks; for example, while pdhB (a high-frequency gene in the unperturbed cradle-to-grave network) is not found to be a high-frequency gene in the perturbed cradle-to-grave network, pdhC *is* among the high-frequency genes of this network, and this trend in gene-type conservation thus produces trends in regulatory path conservation. Such functional conservation further suggests that our network inference process is robust to perturbation at the level of subglobal behavioral trends.

## Discussion

Protein interaction networks and transcriptional regulatory networks represent maps of possible interactions among cellular components, but only a *subset* of these interactions is active at a given time, in a given cell, or under a given condition. In fact, the *dynamic* topology of a cellular network may be radically different from the topology suggested by *static* protein interaction and transcriptional regulatory maps. Our inference process, both by maximizing sparseness in connectivity within the *B. subtilis* networks, as well as by imposing a confidence threshold on interactions, reveals the most salient and active interactions within the network, and thus offers a step towards a dynamic understanding of gene regulation—an understanding that cannot be inferred from systematic mapping methods.

Our graph-theoretical analysis of the topology of gene-regulatory networks, synthesized through a linear correlative process, has demonstrated that the inferred networks not only belong to a larger class of heterogeneous biological graphs, but further, that they produce biologically realistic, and experimentally-verifiable structure at several organizational levels. We characterized our inferred networks as being composed of small, tightly-knit clusters of genes that are connected to one another both through high-degree hub nodes, as well as through other linking pathways between the clusters. We have also demonstrated that the substructures comprising these clusters—the substructures that form the foundations of the networks-- are topologically identical to small-scale motifs that have been observed in other gene-regulatory networks. By examining the functionality of these substructures we have shown not only that the subglobal topologies of our networks both predict and are predicted by the networks' global topologies, but also that the subglobal topologies are highly responsive to environmental stimuli.

While analysis has revealed that our network inference process is not robust at the level of specific regulatory connections, the success of our process in capturing nonrandom trends in regulatory behavior from the global scale to a subglobal scale involving only a handful of nodes suggests that integrating large-scale inference with extensive, multi-scale graph-theoretical analysis offers a powerful technique for constructing networks from highly underdetermined and noisy data. We therefore suggest that the analyses and techniques described in this paper could easily be tailored to the examination of any complex system, and can provide valuable information regarding the characteristic behaviors and features of such systems at multiple scales of complexity.

## Appendix: Pictorial guide to high-significance four and five-node motifs identified in *B. subtilis* cradle-to-grave and amino acid pulse networks

The number beneath each motif is its *Mfinder* software identifier. Table 9 indicates the networks in which each high-frequency motif was found. The table includes only those motifs that were present in two or more networks, and with Z-score> 2.

**Table 6**

| Motif | Cradle-to-Grave 90% | Amino Acid Pulse 90% | Perturbed Cradle-to-Grave 90% | Perturbed Amino Acid Pulse 90% |
|---|---|---|---|---|
| 206 |  | • |  | • |
| 222 | • | • |  |  |
| 2254 |  | • | • | • |
| 2270 | • | • |  |  |
| 926 |  | • |  | • |
| 9118 |  | • | • | • |
| 25502 |  |  | • | • |
| 50088 |  | • |  |  |

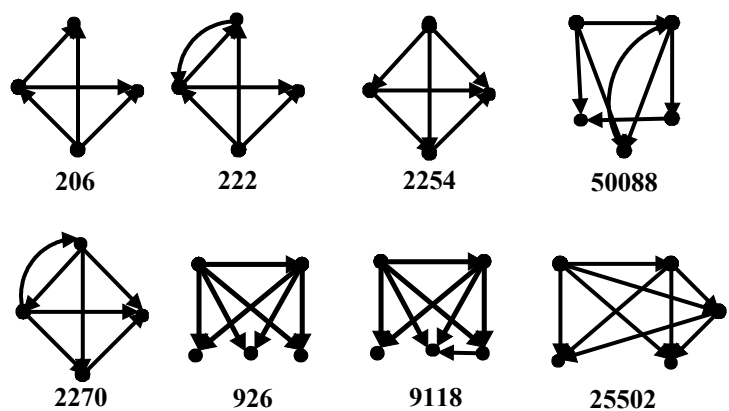

**206**  **222**  **2254**  **50088**

**2270**  **926**  **9118**  **25502**

## Figures and figure captions

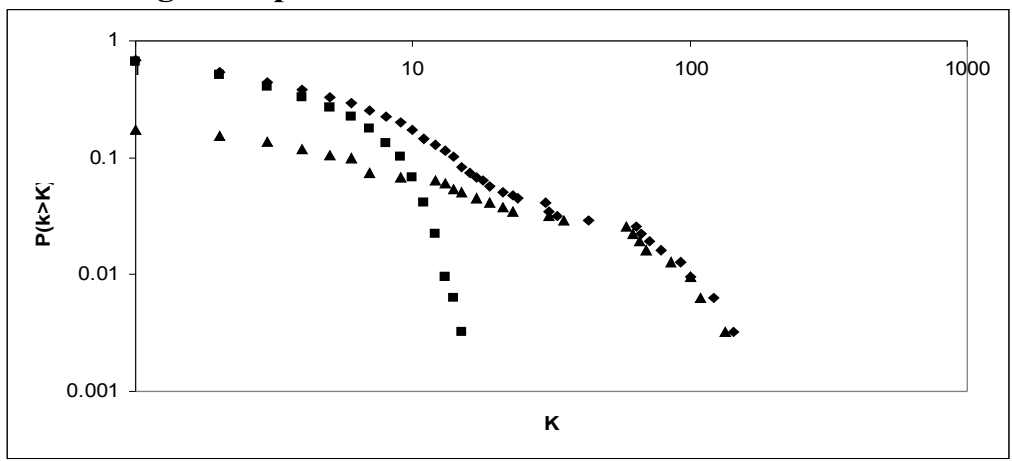

**Fig. 1.** Log-log plot of the cumulative degree distributions for the unperturbed cradle-to-grave network at 90% confidence. Note that the total (diamonds) and out-degree (triangles) distributions are scale-free (with exponential tails), while the in-degree distribution (squares) is more nearly exponential.

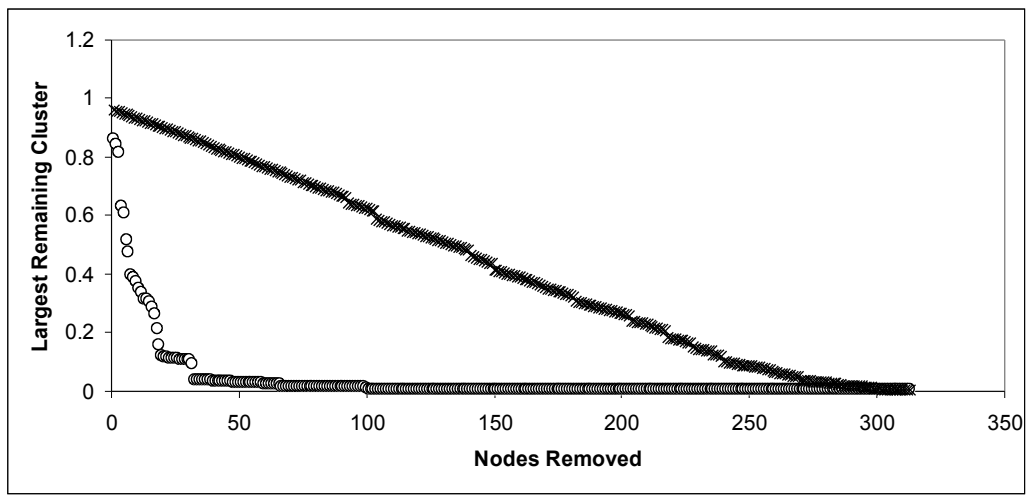

**Fig. 2.** Error-attack plot for unperturbed cradle-to-grave network at 90% confidence: cross hatches represent random removal of nodes, while circles correspond to targeted attack of the network's hubs (nodes whose degree is at least 25% of the maximum degree in the network). The size of the largest remaining cluster decreases approximately linearly when nodes are removed at random, but the network disintegrates rapidly if hubs are targeted for removal. The perturbed network which was inferred in [17] from the same basic data as was used for the unperturbed network, but without first accommodating redundancies in the microarray channel labeling, exhibits similar behavior, and appears approximately equally susceptible to targeted attack.

| **Amino Acid Pulse** | **Cradle-to-Grave** |
|---|---|
| **Feed-forward Loop** 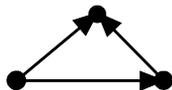 | **Feed-forward Loop with Extra Edge** 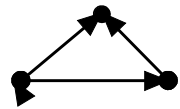 |
| **Double Feed-forward Loop** 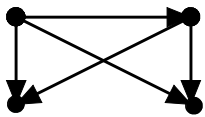 | **Double Feed-forward Loop with Extra Edge** 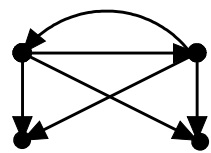 |
| **Triple Feed-forward Loop** 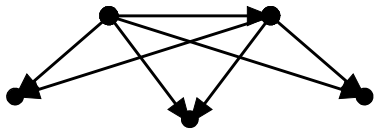 | **Triple Feed-forward Loop with Extra Edge** 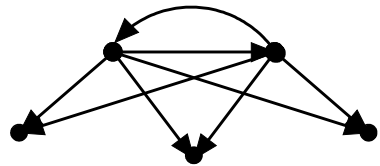 |

**Fig. 3**. Most abundant and statistically-significant motifs in the amino acid pulse and cradle-to-grave networks at 90% confidence.

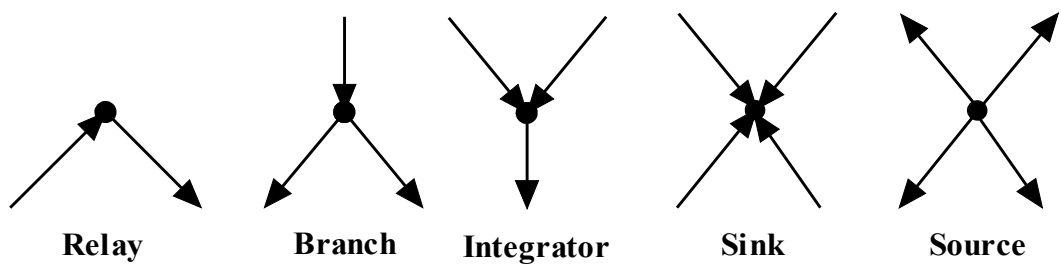

**Fig. 4.** Node roles. Information propagating and processing tasks are conveyed by edge abundance and directionality.

# Tables

**Table 1**. Node and edge composition of each inferred regulatory network at 70%, 80%, and 90% confidence. The networks inferred in reference [17] are denoted as being "perturbed" (see Materials and Methods). The "unperturbed" regulatory interactions reported on in this paper are available from the corresponding author, upon request (e-mail cpc146@phys.psu.edu).

| Network Name and Confidence | Nodes | Edges |
|---|---|---|
| Amino Acid Pulse, 70% | 356 | 1040 |
| Amino Acid Pulse, 80% | 274 | 747 |
| Amino Acid Pulse, 90% | 169 | 405 |
| Perturbed Amino Acid Pulse 70% | 365 | 1058 |
| Perturbed Amino Acid Pulse 80% | 277 | 753 |
| Perturbed Amino Acid Pulse 90% | 154 | 348 |
| Cradle-to-Grave, 70% | 476 | 2722 |
| Cradle-to-Grave, 80% | 396 | 2000 |
| Cradle-to-Grave, 90% | 314 | 1238 |
| Perturbed Cradle-to-Grave 70% | 508 | 2931 |
| Perturbed Cradle-to-Grave 80% | 416 | 2167 |
| Perturbed Cradle-to-Grave 90% | 326 | 1335 |

**Table 2**. Degree ranges and distributions for cradle-to-grave and amino acid data sets at 90% confidence. The in-degree distributions follow exponential functions, $P(k) \sim e^{-\lambda k}$, while the out-degree and total degree distributions follow power laws, $P(k) \sim k^{-\gamma}$.

| Data Set | Degree Range | $1/\lambda$ | $\gamma_{out}$ | $\gamma_{total}$ |
|---|---|---|---|---|
| Amino Acid Pulse 90% | 1-97 | -7.35 | 1.52 | 1.85 |
| Perturbed Amino Acid Pulse 90% | 1-73 | -5.84 | 1.60 | 1.93 |
| Cradle-to-Grave 90% | 1-157 | -8.96 | 1.58 | 1.74 |
| Perturbed Cradle-to-Grave 90% | 1-176 | -8.68 | 1.60 | 1.75 |

**Table 3.** Identity and function of the networks' hubs (nodes whose degree is 25% or more of the highest degree in the network). None of the hubs is a transcriptional regulator.

| Hub Identity | Function | Network |
|---|---|---|
| ypcA | dehydrogenase | Amino Acid Pulse 90% |
| thrB | kinase | Amino Acid Pulse 90% |
| purD | ligase/synthetase | Amino Acid Pulse 90% |
| sdhB | dehydrogenase | Amino Acid Pulse 90% |
| pyrE | phosphoribosyltransferase | Amino Acid Pulse 90% |
| yqhJ | dehydrogenase | Amino Acid Pulse 90% |
| ctaE | oxidase | Cradle-to-grave 90% |
| phrE | activity regulator | Cradle-to-grave 90% |
| atpB | synthase | Cradle-to-grave 90% |
| secE | translocase | Cradle-to-grave 90% |
| sigW | polymerase | Cradle-to-grave 90% |
| argG | synthetase | Cradle-to-grave 90% |
| pdhC | dehydrogenase | Cradle-to-grave 90% |
| ytcF | decarboxylase | Cradle-to-grave 90% |
| gbsA | dehydrogenase | Cradle-to-grave 90% |
| sucC | dehydrogenase | Cradle-to-grave 90% |

**Table 4.** High-frequency genes—i.e. genes that appear in at least two statistically significant motifs-- and their roles in the unperturbed cradle-to-grave and amino acid pulse networks at 90% confidence. Of 26 high-frequency genes in the amino acid pulse network, and 21 high-frequency genes in the cradle-to-grave network, only the five genes that appear in the widest array of high-significance motifs are listed. Their dominant roles are given in bold-faced type. B: branch; Si: sink; So: source; I: integrator; R: relay.

| Network | High-frequency nodes (roles listed under nodes) | | | | |
|---|---|---|---|---|---|
| Cradle-to-Grave | CtaE | phrE | sucC | pdhC | ytcF |
| | **B** | **B** | Si, **B** | I, R, **B** | R, **B** |
| Amino Acid Pulse | SecF | citH | pgi | nadB | purF |
| | **Si** | **Si** | **Si** | **Si** | **Si** |

22**Table 5.** KEGG2 regulatory paths common to both an unperturbed network and its perturbed counterpart. The **Path** column gives the KEGG2 path identifier. The high-frequency genes belonging to these paths are given in columns three and four. While regulatory paths are comprised of different high-frequency genes in the unperturbed and perturbed networks, there is similarity in gene-type between the high-frequency genes in the two varieties of network.

| Path | Function | Constituent Genes in Unperturbed Network | Constituent Genes in Perturbed Network |
|---|---|---|---|
| Cradle-to-Grave | | | |
| *bsu 10* | Carbohydrate/Energy metabolism | pdhB, pgk | gap, pdhC |
| *bsu 190* | Energy metabolism | ctaD, ctaB | ctaE, hpr, atpB |
| *bsu 230* | Nucleotide metabolism | yjbP, hprT, ylcD | adk, yloH, hpr |
| Amino Acid Pulse | | | |
| *bsu 230* | Nucleotide metabolism | adk, guaA, purN, purC, rpoB, yloD, rpoC, purL | purF, purM, yerA |
| *bsu 240* | Nucleotide metabolism | rpoB, rpoC, pyrB | pyrAA, ctrA |
| *bsu 260* | Amino acid metabolism | yqhK, yqhJ, thrS | thrC, thrB |